\documentclass[aps,floatfix,twocolumn,pra,nofootinbib,superscriptaddress,longbibliography]{revtex4-2}
\usepackage{array}
\usepackage{amsmath,amsfonts,amssymb,stackrel}
\usepackage{hyperref}
\hypersetup{colorlinks=true}
\hypersetup{urlcolor=blue}
\hypersetup{citecolor=black}
\hypersetup{menucolor=black}
\hypersetup{linkcolor=black}
\usepackage{cleveref}
\usepackage{float}
\usepackage[T1]{fontenc}
\usepackage[utf8]{inputenc}
\usepackage{tikz}
\PassOptionsToPackage{hyphens}{url}

\newcommand{\be}{\begin{equation}}
\newcommand{\ee}{\end{equation}}
\newcommand{\bea}{\begin{eqnarray}}
\newcommand{\eea}{\end{eqnarray}}

\newcommand{\bb}[1]{\left( #1 \right)}

\newcommand{\bbcro}[1]{\left[ #1 \right]}

\newcommand{\rF}{\textrm{F}}
\newcommand{\ii}{\textrm{i}}
\newcommand{\dd}{\textrm{d}}
\newcommand{\eee}{\textrm{e}}
\newcommand{\rr}{\textbf{r}}

\newcommand{\qq}{\textbf{q}}

\newcommand{\pp}{\textbf{p}}

\newcommand{\kF}{k_{\rm F}}

\newcommand{\TF}{T_{\rm F}}

\newcommand{\pF}{p_{\rm F}}

\newcommand{\upa}{\uparrow}
\newcommand{\dwa}{\downarrow}

\newcommand{\eqqref}[1]{Eq.~\eqref{#1}}
\newcommand{\eqqrefs}[2]{Eqs.~\eqref{#1}--\eqref{#2}}

\usepackage{tikz}
\usetikzlibrary{shapes.geometric, positioning, shadings, arrows.meta}
\usetikzlibrary{calc,patterns,decorations.pathmorphing,decorations.markings}

\tikzset{->-/.style={decoration={
  markings,
  mark=at position .5 with {\arrow{>}}},postaction={decorate}}}
\tikzset{-<-/.style={decoration={
  markings,
  mark=at position .5 with {\arrow{<}}},postaction={decorate}}}
\tikzset{phantom->-/.style={decoration={
  markings,
  mark=at position .5 with {\arrow[scale=2]{>}}},postaction={decorate}}}

\tikzset
  {
    ,my bubble/.style = 
      {
        ,draw=#1!70
        ,fill=#1!10
        ,ellipse
        ,inner sep=1pt
        ,minimum width=2em
        ,minimum height=2em
        ,align=center
      }
    ,my end/.style =
      {
        ,draw=#1!70
        ,top color=#1!10
        ,bottom color=#1!50
        ,minimum height=6em
        ,text width=6em
        ,inner sep=0pt
        ,align=center
      }
    ,my arrow/.style =
      {
        ,>=Stealth
        ,->
        ,draw=black
      }
  }
  
\tikzset{serpent/.style={decoration={snake},postaction={decorate}}}

\begin{document}
\title{Dispersion of first sound in a weakly interacting ultracold Fermi liquid}

\newcommand{\hk}[1]{{#1}}

\author{Thomas Repplinger}
\affiliation{Laboratoire de Physique Théorique,
Université de Toulouse, CNRS, UPS, 31400, Toulouse, France}
\author{Songtao Huang}
\affiliation{Department of Physics, Yale University, New Haven, Connecticut 06520, USA}
\author{Yunpeng Ji}
\affiliation{Department of Physics, Yale University, New Haven, Connecticut 06520, USA}
\author{Nir Navon}
\affiliation{Department of Physics, Yale University, New Haven, Connecticut 06520, USA}
\affiliation{Yale Quantum Institute, Yale University, New Haven, Connecticut 06520, USA}
\author{Hadrien Kurkjian}
\email{hadrien.kurkjian@cnrs.fr}
\affiliation{Laboratoire de Physique Théorique,
Université de Toulouse, CNRS, UPS, 31400, Toulouse, France}
\affiliation{Laboratoire de Physique Théorique de la Matière Condensée,
Sorbonne Université, CNRS, 75005, Paris, France}

\begin{abstract}
At low temperature, a normal gas of unpaired spin-1/2 fermions is one of the cleanest 
realizations of a Fermi liquid. It is described by Landau's theory, where no phenomenological 
parameters are needed as the quasiparticle interaction function can be computed perturbatively
in powers of the scattering length $a$, the sole parameter of the
short-range interparticle interactions. Obtaining an accurate solution of the transport equation nevertheless
requires a careful treatment of the collision kernel, {as the uncontrolled error made by the relaxation time approximations
increases when the temperature $T$ drops below the Fermi temperature}. Here, we study
sound waves in the hydrodynamic regime up to second order
in the Chapman-Enskog's expansion. We find that the frequency $\omega_q$
of the sound wave is shifted above its linear departure as $\omega_q=c_1 q(1+\alpha q^2\tau^2)$ 
where $c_1$ and $q$ are the speed and wavenumber of the sound wave and the typical collision time $\tau$ scales as $1/a^2T^2$.
Besides the shear viscosity, the coefficient $\alpha$ is described by a single
second-order collision time which we compute exactly from an analytical solution of the
transport equation, resulting in a positive dispersion $\alpha>0$. Our results suggest that ultracold atomic Fermi gases are an ideal experimental system
for quantitative tests of second-order hydrodynamics.
\end{abstract}
\maketitle

\section{Introduction}
%\textit{Introduction---}
Landau's Fermi liquid theory is an effective theory which, when it is applicable, greatly simplifies
the description of the many-body dynamics of a system of fermions, breaking it down
to a single kinetic equation on a distribution of dilute quasiparticles.
It is a very successful theory in describing
the phenomenology of a wide class of fermionic systems,
such as liquid ${}^3$He \cite{BaymPethick,Wolfle1990}, electron gases \cite{McClure1966,Behnia2022,Kirkpatrick2022}, quantum gases \cite{Vichi2000},
down to nuclear/neutron matter \cite{Schafer2014,Polls2016}. Nevertheless, if both 
the quasiparticle interaction function and the collision probability \cite{Rainwater1975} are known,
either from measurements \cite{Greywall1986,Okuda1996} or from a microscopic calculation \cite{LandauLipschitzVol9},
Fermi liquid theory provides quantitative predictions on dynamical properties,
such as the transport coefficients \cite{Wilkins1968,Sykes1968,Mohling1976}.

In normal  ${}^3$He, despite decades of research, there is still a discrepancy
between theory and experiment on the value of the transport coefficients. Consider for instance the shear
viscosity $\eta$, which scales with temperature as $T^{-2}$: the theory still underestimates
the product $\eta T^2$ from the measurement by more than
$20\%$ \cite{Wheatley1966,BaymPethick,Reppy1976,Lee2016,Parpia2023}.
This is due to a limited knowledge of the quasiparticle interaction function and collision probability \cite{BaymPethick}, which are not computed
from a microscopic theory, and whose experimental determination is limited to the lowest spherical harmonics.
As a consequence, the exact solutions of
the transport equation \cite{Wilkins1968,Sykes1968,Mohling1976} were never validated experimentally, 
and relaxation time approximations \cite{Khalatnikov1958,BaymPethick} remain in use
today \cite{Lee2016}.

Ultracold gases of fermionic gases provide exciting opportunities to quantitatively test those transport calculations \cite{yaleexp}. 
These gases behave as Fermi liquids 
when the s-wave scattering length $a$ is negative and sufficiently small to open a regime
of temperatures $T_c\leq T\ll T_\rF$ where $T_c$ is the superfluid critical temperature
and $T_\rF$ the Fermi temperature. A Fermi liquid regime may also exist when $a$ is positive and small enough to suppress three-body recombination \cite{Navon2022}. 
The quasiparticle dispersion, interaction function and collision probabilities can be computed perturbatively 
in powers of $k_\rF |a|$ (with $k_\rF$ the Fermi wavenumber) \cite{LandauLipschitzVol9}.
Experimentally, both the interaction strength and the temperature
can be varied such that the typical collision time $\tau$ can 
be adjusted over several orders of magnitude \cite{Grimm2008,Zwierlein2019},
to explore both collisionless and hydrodynamic regimes \cite{yaleexp}.
Flat-bottom potentials \cite{Hadzibabic2013,Hadzibabic2021}, where sound can be excited at very low wavevector $q$
in homogeneous samples, give access to the propagation and attenuation of sound waves in a very controlled environment \cite{Zwierlein2019}.

Theoretically, great efforts were devoted to the calculation of the viscosity at strong coupling, in particular in the unitary
regime $|a|=+\infty$ \cite{Schafer2007,Smith2007,Zwerger2011}, and exact results are available in the high-temperature
virial regime \cite{Schafer2010,Nishida2019,Enss2019,Hofmann2020,Enss2023b}. At intermediate temperatures however,
a controlled approach has not been found due to the absence of a separation of timescales between the collisional and kinetic dynamics \cite{LandauLipschitzVol10sec16}.
The temperature range of the Fermi liquid regime shrinks as the quasiparticle cross section increases with the interaction strength
\cite{Rothstein2022}, and it is eventually hidden by the onset of a superfluid phase at a critical temperature $T_c\approx 0.17T_\rF$ at unitarity.
At temperatures low compared to $T_c$, sound attenuation is dominated by phonon-phonon
interactions \cite{Schafer2007,Zwerger2011,annalen}.

In the weakly interacting normal phase, the transport coefficients
were computed using relaxation time approximations \cite{Vichi2000,Smith2005,Nikuni2009,Enss2023}
(either in the ``variational'' \cite{Smith2005,Enss2023}, or in the original Abrikosov-Khalatnikov 
\cite{Vichi2000,Nikuni2009} formulation), even though their uncontrolled error increases toward low temperatures. 
In this article, we perform an exact calculation of the transport coefficients 
\cite{Wilkins1968,Sykes1968} to lowest order in $k_\rF |a|$, and show that the error
of the relaxation time approximations is significant, up to 25\%.

For negative values of $a$, the quasiparticle interactions 
are attractive \cite{Stringari2010}, which prevents the emergence of a zero sound mode as in liquid ${}^3$He.
We thus lack a parameter similar to $c_0-c_1$, where $c_0$, $c_1$ are the speed
of zero and first sound respectively, to characterize the dispersion of sound, as was done in ${}^3$He \cite{Wheatley1966}.
In this work, we derive the leading-order deviation of the frequency $\omega_q$ from its linear departure $c_1 q$.
To do so, we solve the transport equation to the order $\tau^2$
of the Chapman-Enskog's expansion.
This is the so-called second-order hydrodynamics \cite{ChangUhlenbeck,Uribe2008,Schafer2014secondorder} often used in a relativistic context
to cure the acausality of the diffusion equations characteristic of dissipative hydrodynamics \cite{Israel1976}.
We find the exact solution of the transport equation by decomposing the 3D quasiparticle distribution function
on a basis of orthogonal polynomials adapted to the low-temperature limit \cite{Gran2023}. 
Remarkably, the frequency shift involves only two parameters of the collision kernel: the  viscous relaxation time $\tau_\eta$
and a {second-order viscous time} $t_\eta$, which we both compute exactly.
As for the dispersion of the sound branch \cite{annalen},
we find that $\omega_q$ is above its linear 
departure $c_1 q$, the deviation being proportional to $q^3 \tau^2$ with $\tau$ scaling as $1/a^2T^2$.

\section{Transport equation at low temperature}
%\textit{Transport equation at low temperature---}
Landau's theory postulates that a Fermi liquid is described
by a local quasiparticle distribution $n_\sigma(\pp,\rr,t)$, which is the number
of quasiparticles of spin $\sigma$ having momentum $\pp$ at position $\rr$ and time $t$.
This distribution deviates slightly on average from its value in the
Fermi sea $n^0_{\sigma}(p)=\Theta(p_{\rm F}-p)$ (where $p_{\rm F}$ is the Fermi momentum and $\Theta$ the Heaviside distribution), and
the energy
of an arbitrary quasiparticle configuration is expanded to second
ordrer in $\delta n_{\sigma}^0=n_{\sigma}-n^{0}_{\sigma}$:
\begin{multline}
E=E_0+\sum_{\pp,\rr,\sigma}\epsilon_{\sigma}^0(\pp)\delta n_\sigma^0(\pp,\rr)\\+\frac{1}{2}\sum_{\pp,\sigma,\pp',\sigma',\rr}f_{\sigma\sigma'}(\pp,\pp') \delta n_\sigma^0(\pp,\rr) \delta n_{\sigma'}^0(\pp',\rr)
\end{multline}
%is the fluctuation
%about thermal equilibrium $n_{\sigma}^{\rm eq}(\epsilon_{\rm eq})=1/(1+\eee^{(\epsilon_{\rm eq}-\mu_\sigma)/T})$.
In the general case, the ground state dispersion relation
$\epsilon_\sigma^0$ of quasiparticles and the interaction function $f_{\sigma\sigma'}$
are phenomenological parameters usually reexpressed in terms of 
an effective mass and Landau parameters. In the case of a
weakly interacting gas with contact interactions, these quantities
can be calculated perturbatively in powers of the coupling constant $g=4\pi a/m$ \cite{LandauLipschitzVol9} (we use $\hbar=k_{\rm B}=1$ throughout this work).
To first order in perturbation theory, we have:
\be
\epsilon_{\sigma}^{\rm 0}(\pp)=\frac{p^2}{2m}+g\rho_{\sigma'}, \quad f_{\uparrow\downarrow}={g}/V, \quad  f_{\sigma\sigma}=0
\label{param}
\ee
where $V$ is the volume of the gas, and $\rho_\sigma$ the density of fermions of spin $\sigma$.

The time evolution of the quasiparticle distribution is described by a
transport equation:
\be
\frac{\partial n_\sigma}{\partial t}  +\frac{\partial \epsilon_\sigma}{\partial \pp}  \cdot \frac{\partial n_\sigma}{\partial \rr} - \frac{\partial (\epsilon_\sigma+U_\sigma)}{\partial \rr} \cdot \frac{\partial n_\sigma}{\partial \pp} =I_{\sigma}
\label{Boltzmann1}
\ee
where $U_\sigma$ is an external driving field, $\epsilon_\sigma(\pp,\rr)=\epsilon_{\sigma}^{\rm 0}(\pp)+\sum_{\pp',\sigma'}f_{\sigma\sigma'}(\pp,\pp') \delta n_{\sigma'}^0(\pp',\rr) $ 
is the local energy of the quasiparticles, and the collision integral is given in this weakly interacting
limit by Fermi's golden rule
\begin{widetext}
\begin{multline}
I_{\sigma}(\pp)=\frac{2\pi g^2}{V^2}\sum_{\pp_2,\pp_3,\pp_4} \delta_{\pp+\pp_2,\pp_3+\pp_4}
\delta\bb{\epsilon_\sigma(\pp)+\epsilon_{\sigma'}(\pp_2)-\epsilon_{\sigma}(\pp_3)-\epsilon_{\sigma'}(\pp_4)}
\\
\bbcro{(1-n_\sigma(\pp))(1-n_{\sigma'}(\pp_2))n_{\sigma}(\pp_3)n_{\sigma'}(\pp_4)-n_\sigma(\pp)n_{\sigma'}(\pp_2)(1-n_{\sigma}(\pp_3))(1-n_{\sigma'}(\pp_4))}.
\label{Icoll}
\end{multline}
\end{widetext}
In this expression, all the quasiparticle distributions $n$ are evaluated at position $\rr$ and time $t$, which reflects the assumption that collisions are local and instantaneous.

As we seek the eigenmodes of the transport equation, we assume that the drive is weak,
slowly varying in both time and space,
and we linearize Eq.~\eqref{Boltzmann1} around the thermal equilibrium distribution $n_{\rm eq}(\epsilon_0)=1/(1+\eee^{(\epsilon_0-\mu)/T})$.
We focus here on the unpolarized
case so $\mu$ is the common chemical potential of the two spin species
and $\epsilon_0\equiv\epsilon^0_\upa=\epsilon^0_\dwa$.
We also restrict ourselves to excitations of the total density, and define the total driving field
$U_{\rm tot}=U_\uparrow+U_\downarrow$, that we decompose in Fourier
space $U_{\rm tot}(\rr,t)=\text{Re}(U\sum_\qq \eee^{\ii(\qq\cdot\rr-\omega t)})$.
Restricting to terms of first order in temperature $T$ and the drive intensity
$U$, the transport equation obeyed by $\delta n(\pp)=n_\uparrow(\pp)+n_\downarrow(\pp)-2n_{\rm eq}(\epsilon_0(\pp))$
is, in Fourier space:
\begin{multline}
\bb{\omega-\frac{\pp\cdot\qq}{m}}\delta n(\pp,\qq,\omega)+\frac{\partial n_{\rm eq}}{\partial\epsilon_0}  \frac{\pp\cdot\qq}{m}\bb{g\delta\rho(\qq,\omega)+U} \\ =\ii I_{\rm lin}\label{transport}
\end{multline}
where $\delta\rho(\qq,\omega)$ is the fluctuation of the total density about $\rho_{\rm eq}=\rho_{\uparrow}+\rho_{\downarrow}$ (see \eqqref{deltarho} below),
and the linearized collision integral $I_{\rm lin}$ is given by \eqqref{Ilin} in Appendix.

For $T\ll T_{\rm F}$, transport occurs in a energy shell of typical depth $T$ around the
Fermi energy $\epsilon_{\rm F}$ \cite{NozieresPines1966,BaymPethick}. 
We thus reparametrize the quasiparticle distribution as
\begin{multline}
\delta n(\pp)=U \frac{\partial n_{\rm eq}}{\partial \epsilon_0}\Big\vert_{\epsilon_0=\epsilon_{\rm 0}(\pp)}\nu(\epsilon,\theta)\\  \text{ with } \epsilon=\frac{\epsilon_{\rm 0}(\pp)-\mu}{T}=\frac{p^2/2m-\epsilon_\rF}{ T}
\label{chgmtvar}
\end{multline}
where $\theta$ is the angle between $\pp$ and $\qq$, and we have used the equation of state $\mu=\epsilon_\rF+g\rho_{\rm eq}/2$ of the weakly interacting Fermi gas. At low temperature, the reduced energy $\epsilon$ varies from $-\infty$ to $+\infty$ as the momentum $p$ varies from $0$ to $+\infty$. 

Restricting the transport equation to momenta $p$ lying in the relevant energy shell around the Fermi surface,
we obtain, in the limit $T/T_\rF\to0$ at fixed $\epsilon$ and $\epsilon'$:
\begin{widetext}
\begin{multline}
(c-\cos\theta)\nu(\epsilon,\theta)-\frac{k_\rF a}{\pi}\cos\theta\int_{-\infty}^{+\infty}\dd\epsilon' g(\epsilon')\int_0^\pi\sin\theta'\dd\theta' \nu(\epsilon',\theta')\\
+\frac{\ii}{ \omega_0\tau}\bbcro{\Gamma(\epsilon)\nu(\epsilon,\theta)+\int_{-\infty}^{+\infty}\dd\epsilon' \int \sin\theta'\dd\theta'\frac{\dd\phi'}{2\pi} \mathcal{N}_{\rm od}(\epsilon,\epsilon',u)\nu(\epsilon',\theta')}=-\cos\theta \label{eqtranspo}
\end{multline}
\end{widetext}
where $g(\epsilon)=1/(4\text{cosh}^2(\epsilon/2))$ is the dimensionless density of states, 
$\omega_0=v_\rF q$ is a typical excitation frequency (with $v_\rF=\sqrt{2\epsilon_\rF/m}$ the Fermi velocity), $c=\omega/\omega_0$
is the dimensionless excitation frequency, and
\be
\tau=\frac{\pi}{2ma^2 T^2}
\ee 
is a typical collision time. For numerical applications, we replace $\tau$ by:
\be
\tau_\sigma=\frac{\pi}{2m \sigma T^2}=\frac{\tau}{4\pi}
\ee 
where $\sigma=4\pi a^2$ is the scattering cross section.
This is because the center of the hydrodynamic to collisionless crossover
occurs for $\omega_0\tau_\sigma\simeq 1$ rather than $\omega_0\tau \simeq 1$.
The diagonal part of the collision kernel in \eqqref{eqtranspo} is given by
\be
\Gamma(\epsilon)=\pi^2+\epsilon^2.
\ee
This dimensionless function sets the physical quasiparticle lifetime to $\tau_{\rm qp}(\epsilon)=\tau/\Gamma(\epsilon)$.
The off-diagonal part of the kernel reads
\begin{multline}
 \mathcal{N}_{\rm od}(\epsilon,\epsilon',u)={\frac{\mathcal{S}(\epsilon,-\epsilon')}{\sqrt{2(1+u)}}-2\frac{\mathcal{S}(\epsilon,\epsilon')}{\sqrt{2(1-u)}}}\\
 \text{with }\mathcal{S}(\epsilon,\epsilon')=\frac{\epsilon-\epsilon'}{2}\frac{\text{cosh}\frac{\epsilon}{2}}{\text{cosh}\frac{\epsilon'}{2}\text{sinh}\frac{\epsilon-\epsilon'}{2}}
\end{multline}
where the angular dependence comes through $u=\cos(\pp,\pp')=\cos\theta\cos\theta'+\sin\theta\sin\theta'\cos(\phi-\phi')$.

The conservation of the number of quasiparticle in a collision provides
a relation between $\Gamma$ and $\mathcal{S}$:
\be
\int_{-\infty}^{+\infty} \dd\epsilon' \mathcal{S}(\epsilon,\epsilon')=\int_{-\infty}^{+\infty} \dd\epsilon' \frac{g(\epsilon')}{g(\epsilon)}\mathcal{S}(\epsilon',\epsilon)=\frac{\Gamma(\epsilon)}{2}.
\ee

\section{Hydrodynamic limit}
%\textit{Hydrodynamic limit---}
The hydrodynamic limit is the regime of short collision times
\be
\omega_0\tau\ll 1. \label{hydro}
\ee
In this regime, collisions bring the quasiparticle distribution back to equilibrium
much faster than the typical time $1/\omega_0$ at which the sound wave evolves. Only the few components 
of the distribution that are not affected by collisions (\textit{i.e.} those that belong to the zero-energy
space of the collision kernel) remain significantly excited. 
\eqqref{hydro} defines a dynamical regime, accessible from any equilibrium state
of the phase diagram (i.e. for any value of $a$ and $T$).
Here we have sent $a$ and $T$ to 0 at fixed $\omega_0\tau$
and finally expanded the transport equation in powers of $\omega_0\tau$.
To guarantee that the system remains in the hydrodynamic regime, 
it is required to set $q/\kF=2(T/\TF)^2(\kF a)^2\omega_0\tau_\sigma$ in such a way that $\omega_0\tau_\sigma\ll 1$. 
For a finite-size system (of size $L$), $q$ is limited to $q=\pi/L$ which in typical experiments is $\approx 0.01k_{\rm F}$.

The usual expansion to order $O(\omega_0\tau)^1$ yields the transport
coefficients entering the Fermi-liquid version of the Navier-Stokes equations.
Here, we will push to second-order in the hydrodynamic expansion, that is $O(\omega_0\tau)^2$,
and retrieve the sound dispersion from the quantum analogue of Burnett's hydrodynamics \cite{Uribe2008,Schafer2014secondorder}.

To obtain the exact solution of the transport equation in the hydrodynamic limit,
we expand the distribution $\nu$ over a basis of orthogonal polynomials
\be
\nu(\epsilon,\theta)=\sum_{n,l=0}^{+\infty}\nu_n^l P_l(\cos\theta) Q_n(\epsilon)\label{expanisonnu}
\ee
where $P_l$ are the Legendre polynomials, and $Q_n$
are orthogonal for the scalar product weigthed by the density of states:
\be
\int_{-\infty}^{+\infty} g(\epsilon) Q_n(\epsilon) Q_m(\epsilon)\dd\epsilon =||Q_n||^2\delta_{nm}.
\ee
The polynomials $Q_n$ are obtained by the usual recurrence relation:
\be
\epsilon Q_n=Q_{n+1}+\xi_n Q_{n-1} \quad\text{with}\quad\xi_n\equiv\frac{||Q_n||^2}{||Q_{n-1}||^2}
\ee
We choose $Q_0=1$ and $Q_1=\epsilon$ as the initial condition\footnote{
Note that polynomials $Q_n$ used here differ from the orthogonal polynomials of the momentum $p$
used at nonvanishing temperature \cite{Schafer2010,yaleexp}. The replacement Eq.~\eqref{chgmtvar}
converts even/odd powers of $p^2$ into even/odd powers of $\epsilon$.}.
Note that even and odd polynomials are respectively symmetric
and antisymmetric about the Fermi surface: $Q_n(-\epsilon)=(-1)^nQ_n(\epsilon)$.

By contrast to our exact approach, the relaxation time
approximations \cite{Khalatnikov1958,BaymPethick}
truncate the expansion in Eq.~\eqref{expanisonnu} to $n=0$,
thereby neglecting the energy dependence of the quasiparticle distribution.
The difference between the Abrikosov-Khalatnikov \cite{Khalatnikov1958} 
and the variational formulation \cite{BaymPethick,Smith2005,Schafer2010} lies in the treatment
of the remaining $\epsilon$ dependence of the collision kernel:
Abrikosov and Khalatnikov replaced it by its value in $\epsilon=0$
(in particular they approximate the quasiparticle lifetime by its
value at the Fermi level $\Gamma(\epsilon)\approx\Gamma(0)$), while the variational formulation
averages it over $\epsilon$.

The matrix elements of the collision kernel in the orthogonal  basis $\{Q_n\}$
are given by
\be
\Gamma_{nn'}=\int_{-\infty}^{+\infty} \dd\epsilon g(\epsilon)\Gamma(\epsilon) \frac{Q_n(\epsilon)}{||Q_n||^2} Q_{n'}(\epsilon) \label{Gammannp}
\ee
and
\begin{multline}
\!\!\!\mathcal{N}_{nn'}^l \!=\!\! \int_{-\infty}^{+\infty} \!\!\!\!\!\!\!\!\dd\epsilon \dd\epsilon' g(\epsilon) \int_{-1}^1 \!\!\!\! \dd u P_l(u)  \frac{Q_n(\epsilon)}{||Q_n||^2} \mathcal{N}_{\rm od}(\epsilon,\epsilon',u) Q_{n'}(\epsilon')\label{Nnnp}\\
=\frac{2}{2l+1}\mathcal{S}_{nn'}\bb{(-1)^{l+n'} -2}
\end{multline}
with
\be
\mathcal{S}_{nn'}=\int_{-\infty}^{+\infty} \dd\epsilon \dd\epsilon' g(\epsilon)\mathcal{S}(\epsilon,\epsilon') \frac{Q_n(\epsilon)}{||Q_n||^2} Q_{n'}(\epsilon').
\ee
Note that the subspaces of symmetric and antisymmetric  functions of $\epsilon$ (even and odd $n$ respectively)
 are decoupled: $\Gamma_{nn'}=\mathcal{N}_{nn'}^l=0$ if ${n+n'}$ is odd \cite{Sykes1970}.
 This property is specific to the low-temperature limit where both the density of state 
 $g(\epsilon)$ and energy integration domain are symmetric about the Fermi surface.
Treating separately the odd and even orders,
$\Gamma$ and $\mathcal{S}$ are tridiagonal matrices
and can be expressed analytically as
\bea
\!\Gamma_{nn'}\!\!\!&=&\!\!\!\bb{\pi^2+\xi_{n+1}+\xi_{n}}\delta_{nn'}+\delta_{n-2,n'}+{\delta_{n+2,n'}}\xi_{n+2}\xi_{n+1}\notag\\
\mathcal{S}_{nn'}\!\!\!&=&\!\!\!{2\pi^2\frac{n^2+n-1}{4n^2+4n-3} }\delta_{nn'}\!+\!\frac{\delta_{n-2,n'}}{n(n-1)}\notag
\!+\!\frac{\delta_{n+2,n'}\xi_{n+2}\xi_{n+1}}{(n+2)(n+1)}\\
\eea
with $\xi_n=\frac{\pi^2 n^4}{(2 n + 1)(2 n - 1)}$ as can be shown recursively.

The transport equation projected on the orthogonal basis reads now
\begin{multline}
c\nu_n^l-\bbcro{\frac{l+1}{2l+3}\nu_n^{l+1}+\frac{l}{2l-1}\nu_n^{l-1}}-\frac{2k_\rF a}{\pi} \delta_{l,1}\delta_{n,0}\nu_0^0\\+\frac{\ii}{\omega_0\tau}\sum_{n'}\mathcal{M}_{nn'}^l \nu_{n'}^l=-\delta_{l,1}\delta_{n,0}\label{nunl}
\end{multline}
where we introduce the complete collision tensor
\be
\mathcal{M}_{nn'}^l=\Gamma_{nn'}+\mathcal{N}_{nn'}^l.
\ee
The conservation of the number of quasiparticles, energy, and momentum in
a collision [see Eq.~\eqref{Icoll}] generates zero-energy eigenfunctions of the collision kernel.
In our orthogonal basis, this translates respectively into
$\mathcal{M}_{0n'}^0=\mathcal{M}_{n0}^0=0$,
$\mathcal{M}_{1n'}^0=\mathcal{M}_{n1}^0=0$ and $\mathcal{M}_{0n'}^1=\mathcal{M}_{n0}^1=0$ for all $n,n'\in\mathbb{N}$.
The corresponding equations of motion on the conserved quantities $\nu_0^0$, $\nu_0^1$ and $\nu_1^0$ are
\bea
c\nu_0^0-\frac{\nu_0^1}{3}&=&0, \label{nu00}\\
c\nu_0^1-\bb{1+\frac{2k_\rF a}{\pi}}\nu_0^0-\frac{2}{5}\nu_0^2&=&-1, \label{nu01}\\
c\nu_1^0-\frac{\nu_1^1}{3}&=&0 \label{nu10}.
\eea
To give a physical interpretation to the three conserved components of the distribution $\nu$,
we relate them to the fluctuations 
of the density $\delta\rho$, longitudinal velocity $v_\parallel$ and energy density $e$ of the gas:
\bea
\delta\rho(\qq,\omega)&=&\frac{1}{V}\sum_{\pp} \delta n(\pp,\qq,\omega) \\
\rho_{\rm eq} v_\parallel(\qq,\omega)&=&\frac{1}{V}\sum_\pp \frac{\pp\cdot\qq}{mq}\delta n(\pp,\qq,\omega) \\
\delta e(\qq,\omega)&=&\frac{1}{V}\sum_\pp \bb{\frac{\pp^2}{2m}-\mu}\delta n(\pp,\qq,\omega)
\eea
Remarking that $\nu$ is scaled to the intensity $U$ of the drive (\eqqref{chgmtvar}), one expresses
the conserved components in terms of the linear responses $\chi$ of the system:
\bea 
\nu_0^0&=&-\frac{2\pi^2}{mk_\rF} {\chi_\rho} \text{, }\qquad\chi_\rho\equiv\frac{\delta\rho}{U}\\
\nu_0^1&=&-2k_\rF \chi_v \text{, }\qquad \chi_v\equiv\frac{v_\parallel}{U} \\
\nu_1^0&=&-\frac{6}{mTk_\rF} {\chi_e} \text{, }\qquad \chi_e\equiv\frac{\delta e}{U}
\eea

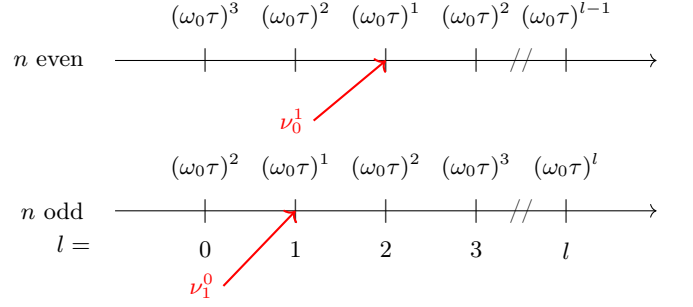
\begin{figure}
\begin{tikzpicture}[xscale=1.2]
\draw[->] (-1,0) node[left=0.3cm] {$n$ odd} node[below left=0.3cm] {$l=$} -- (5,0);
\draw[->] (-1,2) node[left=0.3cm] {$n$ even} -- (5,2);
\draw (0,2) node {|} node[above=0.3cm] {$(\omega_0\tau)^3$} (1,2) node {|} node[above=0.3cm] {$(\omega_0\tau)^2$} (2,2) node {|} node[above=0.3cm] {$(\omega_0\tau)^1$}  (3,2) node {|}  node[above=0.3cm] {$(\omega_0\tau)^2$} (4,2) node {|} node[above=0.3cm] {$(\omega_0\tau)^{l-1}$} ;
\draw (0,0) node {|}  node[above=0.3cm] {$(\omega_0\tau)^2$} node[below=0.3cm] {0} (1,0) node {|}  node[above=0.3cm] {$(\omega_0\tau)^1$} node[below=0.3cm] {1} (2,0) node {|}  node[above=0.3cm] {$(\omega_0\tau)^2$} node[below=0.3cm] {2} (3,0) node {|}  node[above=0.3cm] {$(\omega_0\tau)^3$} node[below=0.3cm] {3} (4,0) node {|}  node[above=0.3cm] {$(\omega_0\tau)^l$} node[below=0.3cm] {$l$};
\draw (3.5,2) node {//};
\draw (3.5,0) node {//};
\draw[->,thick,red] (1.2,1.2) node[left] {$\nu_0^1$} -- (2,2);
\draw[->,thick,red] (0.2,-1) node[left] {$\nu_1^0$} -- (1,0);
\end{tikzpicture}
\caption{\label{fig:echelle} Schematic of the order in $\omega_0\tau$ of the non conserved components $\nu_n^l$ of the quasiparticle distribution.
The red arrows represent the couplings to the conserved quantities $\nu_0^1$ and $\nu_1^0$.}
\end{figure}

In the hydrodynamic limit, only these conserved quantities remain of order unity,
while all the other components pick up one or several factors $\omega_0\tau$.
To evaluate the power of a given component $\nu_n^l$ in $\omega_0\tau$,
one should count the number of projected transport equations \eqref{nunl} needed to reach
a conserved quantity using the couplings appearing in Eq.~\eqref{nunl}, that is, $\nu_n^l\to\nu_n^{l\pm1}$ and, via the collision kernel, $\nu_n^l\to\nu_{n'}^l$ (with $n'$ having the same parity as $n$). 
The components $\nu_{n}^2$ with $n$ even and $\nu_{n}^1$ with $n$ odd, which are directly coupled to the conserved quantities $\nu_0^1$ and $\nu_1^0$ respectively,
are of order $O(\omega_0\tau)^1$. All the other components are subleading, as depicted by Fig.~\ref{fig:echelle}. In particular,
the large $l$ components decay exponentially as $O(\omega_0\tau)^{l-1}$ or $O(\omega_0\tau)^l$, depending on the parity of $n$.

\section{First-sound waves}
%\textit{First sound waves---}
The drive on the right-hand-side of Eq.~\eqref{nunl} is coupled to $\nu_0^1$.
The perturbation it generates is therefore symmetric in $\epsilon$,
that is, $\nu_n^l=0$ for all $n$ odd. In particular the sound wave 
does not generate fluctuations of the energy density: $\nu_1^0=0$.
To leading order in $\omega_0\tau$ the system of Eqs.~\eqref{nu00}--\eqref{nu01} describes an ideal hydrodynamic behavior, that is,
an undamped resonance at $\omega_q=c_1q$ with the first
sound velocity
\be
\frac{c_1}{v_\rF}=\sqrt{\frac{1+\frac{2k_\rF a}{\pi}}{3}}.
\label{c1}
\ee
To study how the resonance deviates from $c_1$,
one must compute the set $\vec{\nu}^2=(\nu_n^2)_{n\in\mathbb{N}}$
of the non-conserved components in the $l=2$ subspace, that we write
as an infinite-dimension vector.
Keeping the leading and subleading terms in Equation \eqref{nunl} for $l=2$, we obtain
\be
\bb{c+\frac{\ii}{\omega_0\tau} \mathcal{M}^2} \vec{\nu}^2=\frac{2}{3}\nu_0^1 \vec{u}_0+O(\omega_0\tau)^2 \label{nu2}
\ee
where we have introduced the matrix $\mathcal{M}^2=(\mathcal{M}_{nn'}^{l=2})_{n,n'\in\mathbb{N}}$ and the unit vectors $(\vec{u}_n)_{n'\in\mathbb{N}}=(\delta_{nn'})_{n'}$.
We have neglected in Eq.~\eqref{nu2} the vector $\vec{\nu}^3$ and the components of $\vec{\nu}^1$
orthogonal to the zero eigenvector $\nu_0^1 \vec{u}_0$:
both are of order $O(\omega_0\tau)^2$ (see Fig.~\ref{fig:echelle}), and hence negligible compared to
the subleading term $c\vec{\nu}^2$ (of order $O(\omega_0\tau)$) in Eq.~\eqref{nu2}.
As is noted in Ref.~\cite{Sykes1970}, the expansion
in powers of $\omega_0\tau$ becomes very tedious
as soon as higher Legendre components have to be taken
into account. Fortunately this is not the case in our calculation to order 
$O(\omega_0\tau)^2$: the transport equation can be truncated
to $l=2$, and only two collision times are needed to express
$\nu_0^2$ from Eq.~\eqref{nu2}:
\be
\nu_0^2=-\frac{2\ii\omega_0 }{3}\nu_0^1\bbcro{\tau_\eta+\ii   \omega_0  ct_\eta^2  } +O(\omega_0\tau)^3
\label{nu02}
\ee
where we introduce the viscous collision time $\tau_\eta/\tau=\vec{u}_0\frac{1}{\mathcal{M}^2}\vec{u}_0$ and
the second-order viscous time $t_\eta/\tau=\sqrt{\vec{u}_0\frac{1}{(\mathcal{M}^2)^2}\vec{u}_0}$.
These two parameters characterize the Burnett hydrodynamical equations on the density and parallel velocity; 
a relaxation time approximation would not distinguish them since it amounts to replacing $\mathcal{M}^2$
by a number. In previous exact calculations, $\tau_\eta$ has been expressed as an infinite serie \cite{Sykes1968,Wilkins1968} or as continued fraction \cite{DasSarma2022}, which converge rather slowly when truncated at $n=n'=n_{\rm max}$: $\tau_\eta-\tau_\eta^{(n_{\rm max})}=O(1/n_{\rm max}^2)$.
Numerically, we obtain
\be
\tau_\eta\simeq 1.079\tau_\eta^{(0)}\simeq 1.288\tau_\sigma \text{\ \  and\ \  }t_\eta\simeq 1.098\tau_\eta^{(0)}\simeq 1.311\tau_\sigma
\ee
where we have introduced the viscous collision time truncated at $n=0$: $\tau_\eta^{(0)}/\tau_\sigma=4\pi/\mathcal{M}^{l=2}_{00}=15/4\pi$.

Plugging Eq.~\eqref{nu02} in Eqs.~\eqref{nu00}--\eqref{nu01}, we obtain a quasi-Lorentzian shape for 
the density response:
\be
\chi_\rho=\frac{mk_\rF}{6\pi^2}\frac{1}{c^2-\bar c_1^2+\frac{4\ii\omega_0 c }{15}\bb{\tau_\eta+\ii \omega_0  c t_\eta^2}}.
\ee
The pole $z_1$ of $\chi_\rho(c)$ is located at:
\be
z_1=\bar c_1-\frac{2\ii}{15}\omega_0\tau_\eta+\frac{2\omega_0^2}{225\bar c_1}(15\bar c_1^2 t_\eta^2-\tau_\eta^2)
\label{z1}
\ee
where $\bar c_1=c_1/v_\rF$.
Going back to the physical units, we write the resonance frequency $\omega_q=\text{Re}(z_1)\omega_0$,
the main result of this work, as
\be
\omega_q\!=\!c_1 q\bbcro{1+\theta(\bar c_1)\epsilon_\rF^2 \tau_\sigma^2\bb{\frac{ q}{mc_1}}^2\!\!\!+O(\omega_0\tau)^3}
\label{omegaq}
\ee
with $\theta(c)\simeq0.917c^2-0.059$. The first deviation from the linear spectrum $c_1 q$ is thus proportional to $q^3$. The dispersion is positive at weak
coupling since $\theta(\bar c_1)\simeq 0.247$ for $\bar c_1= 1/\sqrt{3}$. 
The function $\theta$ vanishes and changes sign at $c_{\rm inv}\simeq0.25$ such that an inversion of the sign of the dispersion may occur in settings where the ratio $c_1/ v_\rF$
is lower than $c_{\rm inv}$.

The damping rate $\Gamma_q=-\text{Im}(z_1)\omega_0$ of the sound wave is determined
only by the shear viscosity $\eta$ in this low-temperature regime:
\be
\Gamma_q=\frac{2}{3m\rho } \eta q^2 \text{ with } \eta=\frac{2}{5}\rho \epsilon_\rF \tau_\eta\simeq 0.5153\rho \epsilon_\rF\tau_\sigma.
\label{gammaq}
\ee
Our value coincides with the exact calculations of $\eta$ done in the context of ${}^3$He \cite{Mohling1976} 
or neutron matter \cite{Polls2016}.
This is a factor $\tau_\eta/{\tau_\eta^{(0)}}\simeq1.08$ above the value in the relaxation time approximation
of Refs.~\cite{Smith2005,Smith2007,Schafer2010} (called the ``variational approximation'' therein).
We note that the underestimation consecutive to the approximation is larger than stated before~\cite{Smith2007,Schafer2010}.
Conversely, in the Abrikosov-Khalatnikov approximation \cite{Khalatnikov1958,NozieresPines1966,Nikuni2009},
the viscosity is overestimated by a factor ${\tau_\eta^{\rm (AK)}}/{\tau_\eta}=4\tau_\eta^{(0)}/3{\tau_\eta}\simeq1.24$.

\section{Thermal conductivity}

Equation \eqref{nu10} on the energy density
allows us to compute the thermal conductivity of the gas.
At low temperature, the energy current $\nu_1^1$ exists only if the
quasiparticle distribution has components odd in $n$, that is if
there is an energy-imbalance across the Fermi surface,
which is not the case in a sound wave.
We compute this energy current by writing down the transport equation for $n$ odd and $l=1$:
\be
c\nu_n^1+\frac{\ii}{\omega_0\tau}\sum_{n'}\mathcal{M}_{nn'}^1\nu_{n'}^1-\nu_n^0-\frac{2}{5}\nu_n^2=0
\ee
To leading order in $\omega_0\tau$, this gives
\be
\frac{\ii}{\omega_0\tau}\mathcal{M}^1 \vec{\nu}_{\rm odd}^1=\nu_1^0 \vec{u}_1+O(\omega_0\tau) \label{bla}
\ee
where $\vec{\nu}_{\rm odd}^1=(0,\nu_1^1,0,\nu_3^1,0,\ldots)$ is the odd component of $\vec{\nu}^1$,
and we recall that $\mathcal{M}$ does not couple the even and odd components. We solve \eqqref{bla}
for $\nu_1^1$, which yields
\be
\nu_1^1=-\ii\omega_0\tau_\kappa\nu_1^0 \quad\text{with}\quad \frac{\tau_\kappa}{\tau_\sigma}=\vec{u}_1\frac{4\pi}{\mathcal{M}^{1}}\vec{u}_1\simeq 0.7478
\ee
Going back to \eqqref{nu10} on the energy density, we obtain the heat diffusion equation
\begin{multline}
\bb{\omega+\ii{D_\kappa} q^2}\nu_1^0=0 \text{ with } D_\kappa=\frac{1}{3} {v_\rF^2} \tau_\kappa \simeq 0.249  {v_\rF^2\tau_\sigma}.
\end{multline}
The thermal conductivity $\kappa$ is related to $D_\kappa$ by $\kappa=c_V D_\kappa$ with
$c_V=m\pF T/3$ the heat capacity.
The relaxation time approximation $\tau_\kappa^{(0)}/\tau_\sigma=15/8\pi$ turns out much worse
for the thermal conductivity than for the viscosity,
with the error reaching $25\%$. We are thus in neat
disagreement with the claim of Ref.~\cite{Schafer2010}
that $|\kappa-\kappa_{(0)}|/\kappa<2\%$.

\section{Conclusion}
%\textit{Conclusion---}
Dispersion of sound was extensively studied in classical gases, where it is very well captured
by second-order (Burnett) hydrodynamics \cite{FochFord}. We have shown here that a
gas of fermions obeys a very similar kind of Burnett hydrodynamics, although
the presence of a Fermi sea leads to a $1/T^2$ behavior of the collision time, as
opposed to $\tau\propto\sqrt{T}$ in a classical gas. A quantum gas is also affected
by a quantum Vlasov force on the left-hand side of the transport equation. 
To leading order in $a$, this force is the gradient of $g\rho(\rr)$, and therefore independent of the quasiparticle momentum
$\pp$. This may not be true to higher order in the interaction strength, and the effect on 
hydrodynamic transport equations remains to be investigated.

Experiments on ultracold gases have often measured  transport
coefficients \cite{Grimm2008,Zwierlein2011Universal,Thomas2011,Kohl2012,Thomas2015,Thomas2024}, 
and thus seem mature to investigate departures from standard Navier-Stokes hydrodynamics.
The dispersion of first sound could be measured 
by tracking the variations of the resonance frequency in the density-density
response function with the excitation wavenumber $q$. Alternatively,
this resonance frequency could be compared to the speed of sound known from the equation of state. 
Dispersive effects also affect the real time dynamics, such as the equilibration after a quench \cite{Enss2024},
or the propagation of shock waves \cite{ChangUhlenbeck,Thomas2007,ondeschoc}.

\begin{acknowledgments}
H.K. acknowledges support from the French Agence Nationale de la Recherche (ANR), under grant ANR-23-ERCS-0005 (project DYFERCO) and the EUR grant NanoX n°ANR-17-EURE-0009 in the framework of the “Programme des Investissements d’Avenir”. N.N acknowledges support by the AFOSR (Grant No. FA9550-23-1-0605), NSF (Grant No. PHY-1945324), DARPA (Grant No. HR00112320038), and the David and Lucile Packard Foundation. T.R. and H.K. thank Yale University for its hospitality.
\end{acknowledgments}

\appendix

\section{Collsion kernel}
We write the linearized collision integral as
\begin{multline}
    I_{\rm lin}=I_{\rm lin,\uparrow}+I_{\rm lin,\downarrow} = -\Gamma(\pp)\delta n(\pp)\\-\frac{1}{L^3}\sum_{\pp'}\bbcro{E(\pp',\pp)-2S(\pp',\pp)}\delta n(\pp') \label{Ilin}
\end{multline}
with the collision kernels $E$ and $S$ (notice that we swap the arguments $\pp\leftrightarrow\pp'$)
\begin{widetext}
\begin{eqnarray}
    E(\pp,\pp') &=& \frac{2\pi g^2}{L^3}\sum_{\pp_3,\pp_4}\delta^{\pp+\pp'}_{\pp_3+\pp_4}\delta\bb{\epsilon_\pp+\epsilon_{\pp'}-\epsilon_{\pp_3}-\epsilon_{\pp_4}}\bbcro{n_{\rm eq}(\epsilon_{\pp'})\overline{n}_{\rm eq}(\epsilon_{\pp_3})\overline{n}_{\rm eq}(\epsilon_{\pp_4})+\overline{n}_{\rm eq}(\epsilon_{\pp'})n_{\rm  eq}(\epsilon_{\pp_3})n_{\rm eq }(\epsilon_{\pp_4})} \label{E}\\
    S(\pp,\pp')&=& \frac{2\pi g^2}{L^3}\sum_{\pp_2,\pp_4}\delta^{\pp+\pp_2}_{\pp'+\pp_4}\delta\bb{\epsilon_\pp+\epsilon_{\pp_2}-\epsilon_{\pp'}-\epsilon_{\pp_4}}\bbcro{n_{\rm eq}(\epsilon_{\pp_2})\overline{n}_{\rm eq}(\epsilon_{\pp'})\overline{n}_{\rm eq}(\epsilon_{\pp_4})+\overline{n}_{\rm eq}(\epsilon_{\pp_2})n_{\rm  eq}(\epsilon_{\pp'})n_{\rm eq }(\epsilon_{\pp_4})} \label{S} 
\end{eqnarray}
\end{widetext}
with $\bar n_{\rm eq}=1-n_{\rm eq}$, and the short-hand notation $\epsilon_{\pp''}$ for $\epsilon_{\rm 0}(\pp'')$. The quasiparticule lifetime $\Gamma$ is related to $E$ and $S$ by $\Gamma(\pp)=\sum_{\pp'}E(\pp,\pp')/L^3=\sum_{\pp'}S(\pp,\pp')/L^3$, and the total collision kernel obeys the conservation laws
\be
\Gamma(\pp)\pp^k+\frac{1}{L^3}\sum_{\pp'}\bbcro{E(\pp',\pp)-2S(\pp',\pp)}\pp'^k=0,\ k=0,1,2
\ee

Using the property of thermal equilibrium
\begin{multline}
n_{\rm eq}(\epsilon_{\pp_1}) n_{\rm eq}(\epsilon_{\pp_2})\overline{n}_{\rm eq}(\epsilon_{\pp_3})\overline{n}_{\rm eq}(\epsilon_{\pp_4})\\=\overline{n}_{\rm eq}(\epsilon_{\pp_1}) \overline{n}_{\rm eq}(\epsilon_{\pp_2}) {n}_{\rm eq}(\epsilon_{\pp_3}) {n}_{\rm eq}(\epsilon_{\pp_4})
\end{multline}
(for four momenta on the energy shell: $\epsilon_{\pp_1}+\epsilon_{\pp_2}=\epsilon_{\pp_3}+\epsilon_{\pp_4}$) together with
the change of variable \eqqref{chgmtvar} and the properties
\bea
E(\pp',\pp) \frac{n_{\rm eq}(\epsilon_{\pp'})\bar n_{\rm eq}(\epsilon_{\pp'})}{n_{\rm eq}(\epsilon_{\pp})\bar n_{\rm eq}(\epsilon_{\pp})}&=&E(\pp,\pp')\\
S(\pp',\pp) \frac{n_{\rm eq}(\epsilon_{\pp'})\bar n_{\rm eq}(\epsilon_{\pp'})}{n_{\rm eq}(\epsilon_{\pp})\bar n_{\rm eq}(\epsilon_{\pp})}&=&S(\pp,\pp')
\eea
\be
T\frac{\partial n_{\rm eq}}{\partial \epsilon_0}\Big\vert_{\epsilon_0=\epsilon_\pp}=-n_{\rm eq}(\epsilon_{\pp})(1-n_{\rm eq}(\epsilon_{\pp}))
\ee
we swap the arguments of $E$ and $S$ as we move the transport equation \eqqref{transport}
from the distribution $\delta n$ to $\nu$:
\begin{multline}
\bb{\omega-\frac{\pp\cdot\qq}{m}}\nu (\pp)+ \frac{\pp\cdot\qq}{m}\bb{g\delta\rho+U}\\=-\ii \bb{ \Gamma(\pp) \nu(\pp)+\frac{1}{L^3}\sum_{\pp'}\bbcro{E(\pp,\pp')-2S(\pp,\pp')} \nu(\pp')}
\end{multline}
Note that in terms of $\nu$, the fluctuations of the total density are $\delta\rho=(1/V)\sum_{\pp'} ({\partial n_{\rm eq}}/{\partial \epsilon_{\rm 0}})_{\epsilon_{\rm 0}=\epsilon_{\pp'}} \nu(\pp')$.

In \eqqrefs{E}{S}, we eliminate $\pp_4$ using momentum conservation, and
convert the sums into integrals  using 
\be
\frac{1}{L^3}\sum_{\pp_i}\to\frac{mT\pF}{(2\pi)^2}\int_{-\infty}^{+\infty}\dd\epsilon_i\int_0^\pi \sin\vartheta_i\dd\vartheta_i\int_0^{2\pi} \frac{\dd\varphi_i}{2\pi} \label{conversion}
\ee
where $i=2,3$ for $E$ and $S$ respectively, $\epsilon_i=(\epsilon_{\rm 0}(\pp_i)-\mu)/T$, $\theta_i$ is the polar angle between $\pp_3$ and $\pp+\pp'$ or $\pp_2$ and $\pp-\pp'$.
Such choices of spherical frames, together with the isotropic collision probability $g^2$, guarantees that the integrand is independant of the azimuthal angles $\varphi_i$.
Note that the conversion \eqqref{conversion} approximates $p_i^2\dd p_i\simeq m T p_F\dd \epsilon_i$,
which is valid up to corrections in $O(T)$. To leading order in $T$, the energy-conservation constraints can be
replaced by their $T=0$ version, namely
\bea
\!\!\!\!\!\!\!\!\!\epsilon_\pp+\epsilon_{\pp'}-\epsilon_{\pp_3}-\epsilon_{\pp_4}\!\!\!&=&\!\!\!\frac{2\pF^2}{m}\bbcro{\cos\frac{\alpha}{2}\bb{\cos\theta_3\!-\!\cos\frac{\alpha}{2}}}\\
\!\!\!\!\!\!\!\!\!\epsilon_\pp+\epsilon_{\pp_2}-\epsilon_{\pp'}-\epsilon_{\pp_4}\!\!\!&=&\!\!\!\frac{2\pF^2}{m}\bbcro{\sin\frac{\alpha}{2}\bb{\cos\theta_2\!-\!\sin\frac{\alpha}{2}}}
\eea
The Dirac function can then be used to perform the remaining angular integration over $\theta_i$. There remains to integrate
over the reduced energy $\epsilon_i$, using:
\begin{widetext}
\bea
\int_{-\infty}^{\infty} \dd \epsilon_{3} \bbcro{n(\epsilon')\bb{1-n(\epsilon_3)-n(\epsilon+\epsilon'-\epsilon_3)}+n(\epsilon_3)n(\epsilon+\epsilon'-\epsilon_3)} &=& \mathcal{S}(\epsilon,-\epsilon')\\
\int_{-\infty}^{\infty} \dd \epsilon_{2} \bbcro{n(\epsilon_2)\bb{1-n(\epsilon')-n(\epsilon+\epsilon_2-\epsilon')}+n(\epsilon')n(\epsilon+\epsilon_2-\epsilon')} &=& \mathcal{S}(\epsilon,\epsilon')
\eea
\end{widetext}
The result are collision kernels whose angular and energy dependance are factorized:
\bea
E(\pp,\pp')&=& \frac{m^2T}{2\pi p_{\rm F}}\frac{g^2}{2\cos(\theta/2)}\mathcal{S}(\epsilon,-\epsilon')\\
S(\pp,\pp')&=& \frac{m^2T}{2\pi p_{\rm F}}\frac{g^2}{2\sin(\theta/2)}\mathcal{S}(\epsilon,\epsilon')
\eea
This remarkable property is specific to Fermi liquids and distinguishes them
from classical gases. It is a direct consequence of the freezing of the collisions to the vicinity of the
Fermi surface when $T\to0$.

To derive the transport equation  in the low-temperature limit \eqqref{eqtranspo},
there remains to divide \eqqref{transport} by $\omega_0(\partial n_{\rm eq}/\partial \epsilon_0)_{\epsilon_0=\epsilon_0(\pp)}$
and expand each term in powers of $T$ at fixed $\epsilon$ and $\epsilon'$ (which amounts to replacing $p$ and $p'$ by $\pF$
outside functions that are peaked about the Fermi sphere).

\section{Observability of the spectrum in the density response function}

\begin{figure}
\includegraphics[width=0.99\columnwidth]{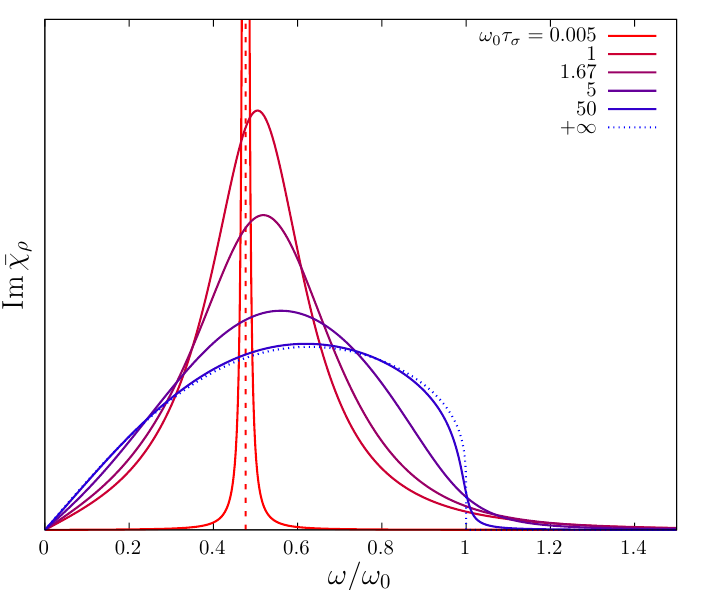}
\caption{\label{transition}
The transition from hydrodynamic (red curves) to collisionless regime (blue curves)
in the density response
of a weakly interacting Fermi gas with $k_\rF a=-0.5$.
The broadening and the positive shift of the resonance frequency from the first sound
velocity (vertical dashed line, Eq.~\eqref{c1}) is clearly visible at small $\omega_0\tau_\sigma$.
The curve are obtained by numerically solving Eq.~\eqref{nunl} truncated to $n_{\rm max}=100$
and $l_{\rm max}$ ranging from 20 in the hydrodynamic regime to 8500 in the collisionless regime.}
\end{figure}

In this section, we verify that the attenuation and dispersion coefficients appearing in Eqs.~\eqref{omegaq}--\eqref{gammaq}
can be accurately recovered from the density response function $\bar\chi_\rho(c)\equiv\nu_0^0(c)$, which is the main
experimental observable for the propagation of sound \cite{yaleexp,Zwierlein2019,Vale2020}.
To do so, we solve Eq.~\eqref{nunl} numerically by truncating it to $n_{\rm max}=100$
and $l_{\rm max}$ ranging from 20 in the hydrodynamic regime to 8500 in the collisionless regime.
Examples of numerically computed spectra are shown in Fig.~\ref{transition};
we observe that the shift $\omega_q-c_1 q$ is positive everywhere in the hydrodynamic
to collisionless crossover.

On the hydrodynamic side of the crossover ($\omega_0\tau_\sigma<0.5$), and for $k_\rF a=0$ ($c_1=v_\rF/\sqrt{3}$)
we fit $\text{Im}\bar\chi_\rho(c+\ii0^+)$
to a Lorentzian function:
\be
f_{\rho}^{\rm fit}(c)= \frac{Z}{(c-\bar c_1-B\omega_0^2 \tau_\sigma^2 )^2+A^2\omega_0^2 \tau_\sigma^2}
\label{lorentzienne}
\ee
where $A,B$ and $Z$ are real fitting parameters.

\begin{figure}
\includegraphics[width=0.99\columnwidth]{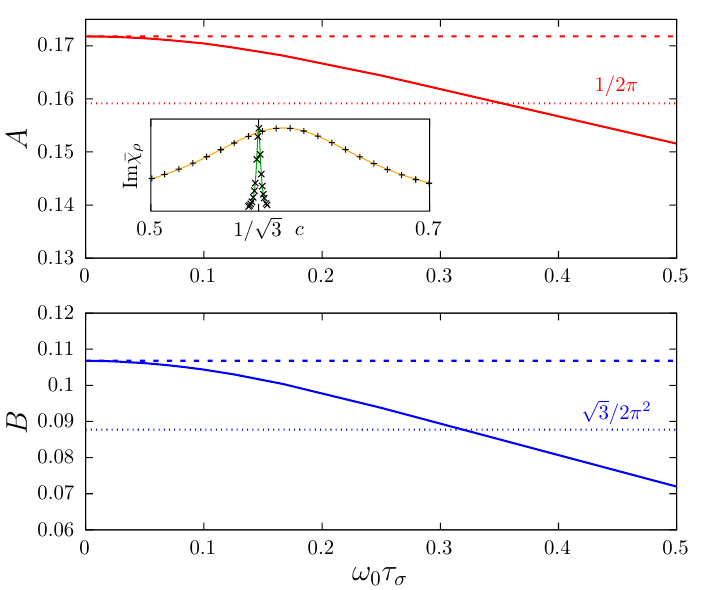}
\caption{\label{gammac}
The sound attenuation and dispersion coefficients $\gamma$ and $\Delta c$ 
obtained by fitting a Lorentzian function
Eq.~\eqref{lorentzienne}  (green and orange lines in inset, respectively for $\omega_0\tau_\sigma=0.01$ and $\omega_0\tau_\sigma=0.5$) 
to the density response $\bar\chi_\rho$ obtained by numerically solving Eq.~\eqref{nunl} at non vanishing $\omega_0\tau_\sigma$ (symbols in the inset).
The horizontal dashed and dotted line show respectively the 
exact hydrodynamic limits Eqs.~\eqref{limhydro1} and
\eqref{limhydro2},
and the relaxation time approximation ($\tau_\eta,t_{\eta}\to\tau_\eta^{(0)}$).
This figure is drawn in the limit $k_\rF a\to 0$ ($c_1/v_\rF=1/\sqrt{3}$).}
\end{figure}

Fig.~\ref{gammac} compares the fitted values at non-vanishing $\omega_0\tau$
to the expected limit when $\omega_0\tau\to0$, revealing a very good convergence to
\bea
B&\underset{\omega_0\tau\to0}{\to}&\frac{\textrm{Re}(z_1-\bar c_1)}{(\omega_0 \tau_\sigma)^2}=\frac{2\sqrt{3}}{225}\frac{5 t_\eta^2-\tau_\eta^2}{\tau_\sigma^2} \label{limhydro1}\\
A&\underset{\omega_0\tau\to0}{\to}&-\frac{\textrm{Im}(z_1)}{\omega_0 \tau_\sigma} =\frac{2}{15}\frac{\tau_\eta}{\tau_\sigma} \label{limhydro2}
\eea

To estimate the order of magnitude of the second-order hydrodynamical effects in experiments,
we use the parameters $k_{\rm F}a=-0.67$, $T/T_{\rm F}=0.17$, 
$q/k_{\rm F}=0.01$ taken from Ref.~\cite{yaleexp} (see Fig.~4 therein).
This corresponds to $\omega_0\tau_\sigma\simeq0.42$ for which
a first sound resonance is still clearly visible in Fig.~4a of Ref.~\cite{yaleexp}. Plugging these experimental parameters
in \eqqref{z1}, we estimate the relative frequency shift
and damping rate to be
\bea
\frac{\omega_q}{c_1q} -1&\simeq& 3\% \\
\frac{\Gamma_q}{c_1q} &\simeq&13\%
\eea
\hk{The small value of $\omega_q-c_1 q$ is challenging to measure, but it could be within experimental reach in the near future.
This would allow to distinguish between supersonic  $\omega_q>c_1 q$ and
subsonic  $\omega_q<c_1 q$ dispersion.}
%With an uncertainty on the measurement of $\chi_\rho$ corresponding to state-of-the-art experiments,
%the precise value of $\omega_q-c_1 q$
%will likely be hidden by the broadening of the response function,
%which is the dominant effect. 

%
%A Fermi liquid, just like a nearly ideal gas (at arbitrary temperatures)
%has no bulk viscosity \cite{BaymPethick,NozieresPines1966}. We attribute
%this to effective quadratic behavior of the dispersion relation
%$\epsilon_\pp-\mu\simeq (p^2-\pF^2)/2m$. After decomposition
%onto the $Q_n$ polynomials, this properties decouples the components
%$\nu_{n>0}^0$ from the conserved $\nu_0^1$, thereby removing
%a dissipative term in the velocity equation.
%Finally the spin diffusivity is outside the scope of this article,
%which did not consider fluctuations of the polarization $\delta n_\upa-\delta n_\dwa$.


\bibliography{/Users/hkurkjian/Documents/Latex/HKLatex/biblio}

\end{document}